# EFFECT OF THE SUPERCONDUCTING WIGGLER ON THE DELSY BEAM DYNAMICS

*P. Beloshitsky[1,2], E. Levichev[3], I. Meshkov[1], I. Titkova[1]*
[1] - Joint Institute for Nuclear Research, Dubna, Russia
[2] - CERN, Geneva, Switzerland
[3] - Budker Institute of Nuclear Physics, Novosibirsk, Russia

## 1. Abstract

The project DELSY is being under development at JINR, Dubna, Russia. This synchrotron radiation source [1-3] is dedicated to the investigation on condensed matter physics, atomic physics, biology, medicine, chemistry, micromechanics, lithography and others. The storage ring DELSY is an electron storage ring with the beam energy 1.2 GeV and 4 straight sections to accommodate accelerator equipment and insertion devices. One of the straight sections is intended for a 10 T superconducting wiggler (wavelength shifter) and one for the undulator with 150 periods and a magnetic field of 0.75 T. The wiggler will influence many aspects of beam dynamics: linear motion, dynamic aperture, emittance, damping times etc. The problem is rather serious for the DELSY machine because the energy of the electron beam is small while the wiggler's magnetic field is strong.

In this paper we consider two models of the wiggler's magnetic field with and without the focusing caused by the sextupolar field of the wiggler as we need to develop the requirements to the wiggler design. We study the influence of the 10 T wiggler on the beam dynamics in the DELSY storage ring and propose a possible scheme to cure it [2-4]. The combined work of the insertion devices is presented too.

## 2. Wiggler model

Since the magnetic field of the wiggler has variation in both transverse and longitudinal direction, it is not easy to represent it by a step-wise function as is usually done with dipole and quadrupole magnets in a hard edge approximation approach. Basically, investigation of the charged particle motion in the wiggler field is performed by special tracking routines which use some of canonical integration techniques. We shall elaborate a simple but still realistic wiggler model to use it in a typical accelerator simulation codes.

For this purpose we use the magnetic measurement data of the 10 T wiggler that was produced by BINP for the Spring-8 synchrotron light source [5]. The result of the step-by-step mapping of the wiggler's magnetic field is fitted by the spline approximation. For this approximation, the on-axis field expansion coefficients, the trajectory and the angle deviation were found numerically (Fig.1, 2).

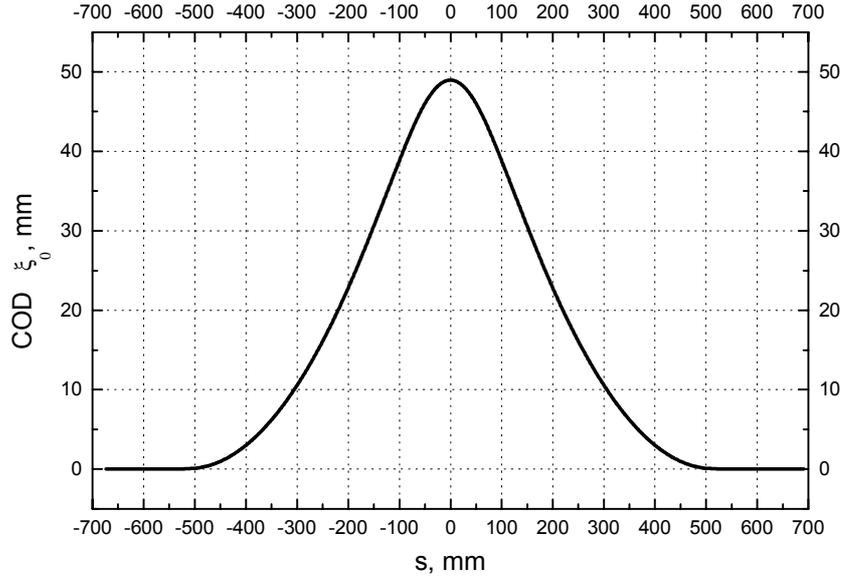

Fig.1. Horizontal closed orbit deviation in the 10 T wiggler

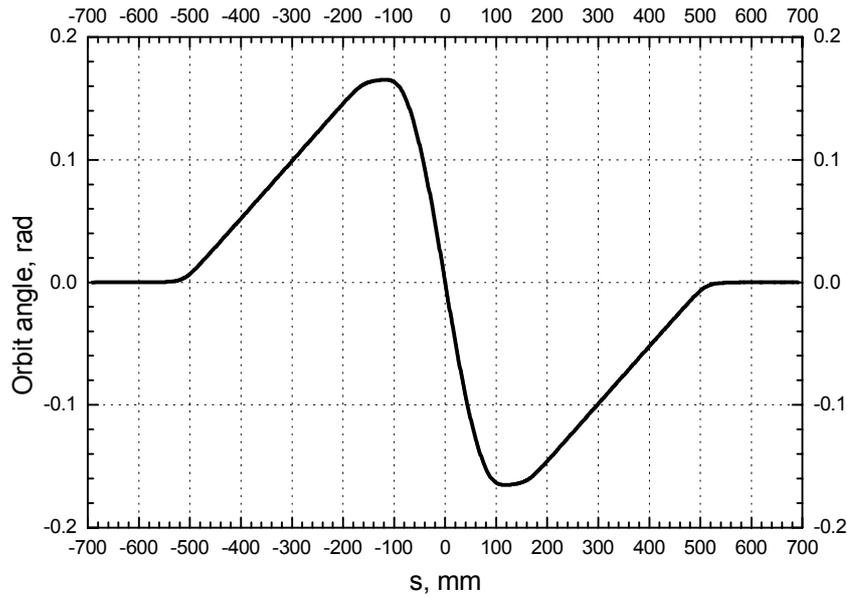

Fig.2. Orbit angle in the 10 T wiggler

Knowing the beam orbit we need to write down equations of particle motion in the vicinity of the orbit in the curvilinear coordinate system that relates to this orbit. The magnetic field expansion up to the octupole order is given below:

$$B_z = B_{z00} + B_{z10}x + B_{z20}x^2 + B_{z02}z^2 + B_{z30}x^3 + B_{z12}xz^2;$$
$$B_x = B_{x01}z + B_{x11}xz + B_{x03}z^3 + B_{x21}x^2z,$$

(1)

where

$$B_{z00} = a + \frac{1}{2}b\xi_0^2; \quad B_{z10} = b\xi_0 - a'\theta - \frac{1}{2}b'\xi_0^2\theta;$$

$$B_{x01} = b\xi_0 - a'\theta - \frac{1}{2}b'\xi_0^2\theta; \quad B_{z20} = \frac{1}{2}b - b'\xi_0\theta$$

$$B_{x11} = b - 2b'\theta\xi_0; \quad B_{z02} = -\frac{1}{2}\left(b + a'' + \frac{1}{2}b''\xi_0\right);$$

$$B_{z30} = -\frac{1}{2}b'\theta; \quad B_{x03} = \frac{1}{6}(b'\theta - b''\xi_0 + a'''\theta);$$

$$B_{z12} = \frac{1}{2}(b'\theta - b''\xi_0 + a'''\theta); \quad B_{x21} = -\frac{3}{2}b'\theta.$$

In the above expressions $\xi_0$ and $\theta$ are the reference orbit in the wiggler (Fig.1) and the angle deviation (Fig.2), $a$ and $b$ are the dipole and sextupole field expansion coefficients in the Cartesian frame that refers to the wiggler axis. The coefficients $a$ and $b$ have been extracted from the wiggler magnetic mapping (Fig.3, 4).

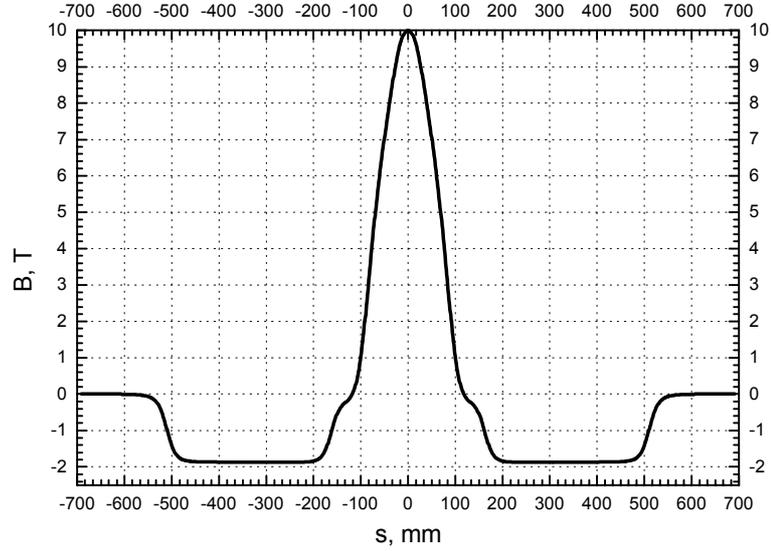

Fig.3. Wiggler dipole field $a$ as a function of the longitudinal coordinate

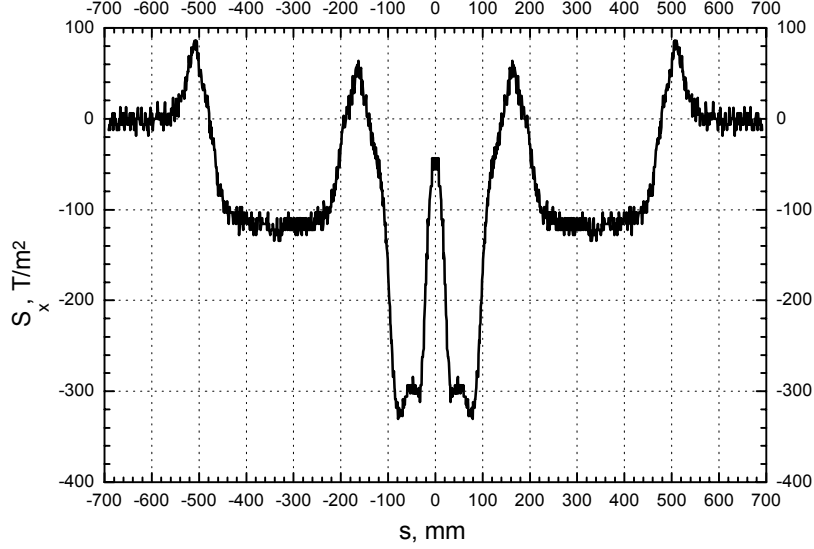

Fig.4. Transverse sextupole component *b* as a function of the longitudinal coordinate

The field expansion (1) is rather complex to be used in typical accelerator codes and we have to study each term in the above expressions separately to simplify the wiggler field representation. During this analysis we shall take into consideration the fact that the betatron functions are practically constant along the wiggler length. This fact allows us to use averaged values instead of actual longitudinal distributions. The wiggler field is symmetric in three planes, therefore only odd-order field components exist in the horizontal direction in medial plane. Even components of the magnetic field may appear in the case of orbit distortion inside the wiggler and can be calculated by transforming the coordinate system. For the wiggler of Spring-8 the dipole and sextupole components have been measured. On the basis of these data the quadrupole component was calculated. To calculate the octupole component, the next odd-order field component is required, but the decapole component has not been measured. Actually, the main acting nonlinearity is sextupole. For this reason we use model the field expansion (1) up to the sextupole order for the wiggler's magnetic field.

*2.1. Linear model of the wiggler field*

The simplest way to construct the model is to make it from the set of hard-edge magnets with the field integrals equal to the measured ones. But this model has strong drawback: the focusing properties of the wiggler (which appear due to presence of the nonlinear magnetic field and big orbit offset and angle in wiggler) will not be taken into account.

The dipole field at the wiggler orbit is described by the expansion coefficients $B_{z00}$ in (1):

$$B_{z00} = a + \frac{1}{2}b\xi_0^2 . \qquad (2)$$

The first term in (2) corresponds to the on-axis dipole field and the second one describes the sextupole contribution, which is rather small in comparison with the pure dipole field. The central region field integral evaluated numerically from the measurement for the region (-119, 119) mm (Fig.3) is equal to

$$I_B = \int_{-119}^{119} B ds = 1{,}392 \text{ T·m} . \qquad (3)$$

In this case the bending angle of the central pole of the wiggler $\alpha_w$ that is determined by the field integral value is

$$\alpha_w = \frac{\int_{-119}^{119} B ds}{B\rho} = 0{,}3305 \text{ rad.} \qquad (4)$$

The wiggler focusing is described by the expansion coefficients $B_{z10}$, $B_{x10}$ in (1):

$$B_{z10} = B_{x10} = b\xi_0 - a'\theta - \frac{1}{2}b'\xi_0^2\theta . \qquad (5)$$

The first term is due to the transverse variation of the magnetic field, the second is a "dipole edge" focusing (here $a' = \frac{da}{ds}$ is the longitudinal derivative of the dipole field) and the third is the function of the particle orbit, longitudinal derivative of the sextupole coefficient, etc. The edge focusing provides the major effect, the data obtained for the edge focusing ($G_e = -\frac{dB}{ds}\theta$ T/m) are presented in Fig.5.

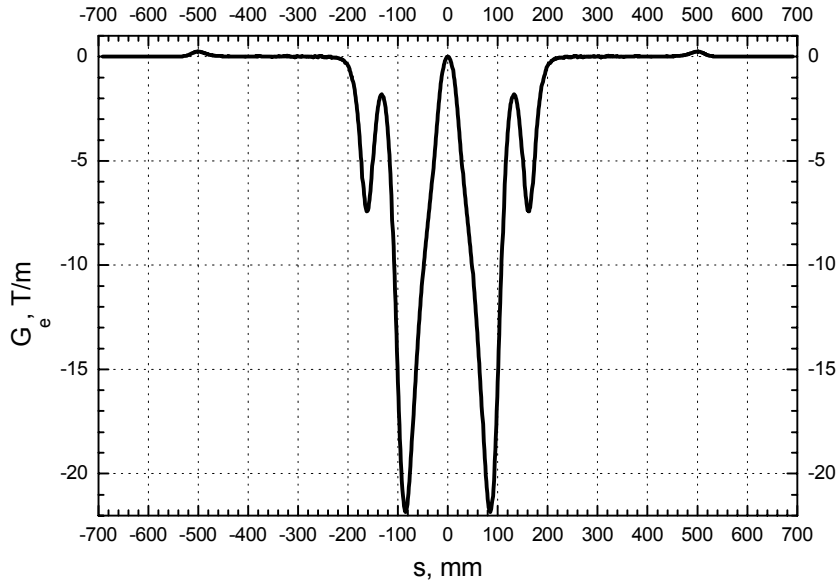

Fig.5. Edge focusing in the wiggler

From these data with the aid of the numerical evaluation we can calculate the vertical focusing of the central pole of the wiggler $I_{zw}$:

$$I_{zw} = \int_{-\frac{L_w}{2}}^{\frac{L_w}{2}} k_z ds = -\frac{1}{B\rho} \int_{-\frac{L_w}{2}}^{\frac{L_w}{2}} \frac{dB}{ds} x' ds = 0{,}776 \text{ m}^{-1}. \qquad (6)$$

From the data of Fig.3 we can see that

$$\int_{-\frac{L_w}{2}}^{\frac{L_w}{2}} \frac{1}{\rho^2} ds = \int_{-\frac{L_w}{2}}^{\frac{L_w}{2}} \left(\frac{B(s)}{B\rho}\right)^2 ds = 0{,}777 \text{ m}^{-1}.$$

As $\frac{1}{\rho^2} = k_x + k_z$, where $k_x$, $k_z$ are the coefficients of the horizontal and vertical focusing, it is clear that the negative portion of the $k_x$ compensates the positive one, so the resulting value for the horizontal plane should be small in comparison with the vertical focusing. This represents the well-known fact: if there is no horizontal variation of the wiggler's magnetic field (sextupole component is equal to zero), the wiggler focuses the beam only in the vertical plane, while horizontal focusing is exactly zero.

Now, when we have prepared field and focusing integrals from the results of magnetic measurements, we can start the construction of the linear model of the wiggler magnetic field. It is evident that more serious attention should be paid to the central pole with meeting the following requirements:

1) conservation of the pole bending angle $\alpha_w$ (4)

2) conservation of the integral horizontal focusing $I_{xw} = \int_{-\frac{L_w}{2}}^{\frac{L_w}{2}} \left( \frac{1}{\rho^2(s)} - k_z(s) \right) ds$

3) correct representation of the wiggler radiation properties according to the so-called "fourth synchrotron radiation integral". This integral is taken over the bending field region and is defined as

$$I_4 = \int_{BM} \frac{(1 - 2n(s))D(s)}{\rho^3(s)} ds \qquad (7)$$

where $D$ is the dispersion function and $n$ is the gradient field index. The central pole of the wiggler has side regions where the dipole field is combined with the gradient field (edge focusing). These regions correspond to the second term in (7) and can influence on the damping partition numbers, energy spread, horizontal emittance, etc.

We describe the central pole of the wiggler by a set of 3 sector dipoles as is shown schematically in Fig.6. BM1 corresponds to the central region of the pole field and focuses the beam only due to $\rho^{-2}$ but has no vertical focusing. Two dipoles BM2 correspond to the side region of the wiggler pole with both the dipole and the gradient field to represent (7) properly. Addition of the side dipoles in the model improves precision of the second and third synchrotron radiation integrals $I_2$ and $I_3$ as well.

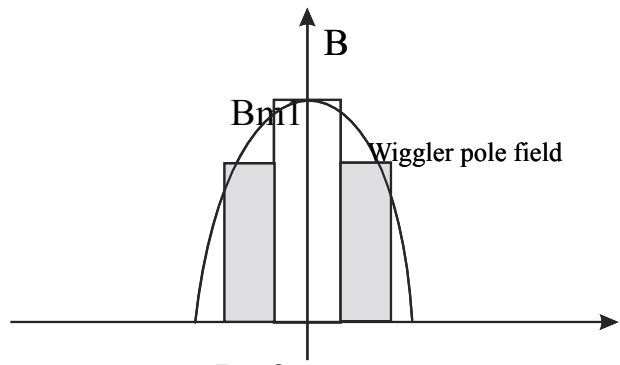

Fig.6. Model of the central pole of the wiggler

The above requirements are written as

$$\begin{cases} \alpha_w = \dfrac{L_1}{\rho_1} + 2\dfrac{L_2}{\rho_2} \\ I_{xw} = \dfrac{L_1}{\rho_1^2} - k_{z1}L_1 + 2\dfrac{L_2}{\rho_2^2} - 2k_{z2}L_2 \\ I_{zw} = k_{z1}L_1 + 2k_{z2}L_2 \end{cases} \qquad (8)$$

Here $L$, $\rho$ and $k$ are the length, bending radius and vertical and horizontal focusing strength of the relevant dipole magnet. To solve these equations we fix the following parameters: $L_1$=0.06 m (compromise between the length of the vertical and horizontal focusing distribution in the actual wiggler field), $\rho_1=\rho_w$=0.4 m at 1.2 GeV. With this choice solution of (8) yields the following parameters of the model for the central pole of the wiggler:
BM1: $L_1$=0.06 m, $\rho_1$=0.4 m, $k_{z1}$=0 m$^{-2}$,
BM2: $L_2$=0.0406 m, $\rho_2$=0.45 m, $k_{z2}$=9.551 m$^{-2}$.

The requirements to the side poles are weaker in comparison with those to the central one, so we can write them in the same way but with a single dipole magnet. We set the distance between central and side poles 0.049 m and the efficient length of the side pole $L_s$=0.35 m (Fig.3). From the side pole bending angle $\alpha_{ws}=\alpha_w/2$ we can calculate the bending radius: $\rho_s$=2.118 m. The side pole focusing occurs in a usual way as for any rectangular magnet, because the sextupole component on the beam orbit is small for the side pole (Fig.4). In our case the focusing strength of the side pole is $k_{zs} = \dfrac{1}{\rho_s^2} = 0.2229$ m$^{-2}$.

*2.2. Nonlinear wiggler fields*

According to (1) the wiggler produces sextupole-like and octupole-like non-linear magnetic fields. The sextupole-like non-linearity is given in (1) by three coefficients $B_{z20}$, $B_{z02}$, $B_{x11}$, but the analysis shows that only the actual sextupole $b(s)$ should be taken into account. The edge field contribution $a''$ is large (Fig.7) but its integral is exactly equal to zero. This well-known fact can be shown with explicit calculation of the integral:

$$\int \dfrac{d^2 a}{ds^2} ds = \left.\dfrac{da}{ds}\right|_{-\infty}^{+\infty} = 0.$$

Other terms in $B_{z20}$, $B_{z02}$ and $B_{x11}$ depend on $\xi_0^2$ or $\xi_0\theta$, hence are very small. Therefore, for our model we can use only the actual sextupole distribution $b(s)$ that is plotted in Fig.4. According to this plot we propose to represent the wiggler sextupoles as a set of two nonlinear kicks with the integral strength $\displaystyle\int_0^{L_w} k_{sex}ds = \dfrac{L_w}{2B\rho}\dfrac{d^2 B}{ds^2} = -12{,}95$ m$^{-2}$.

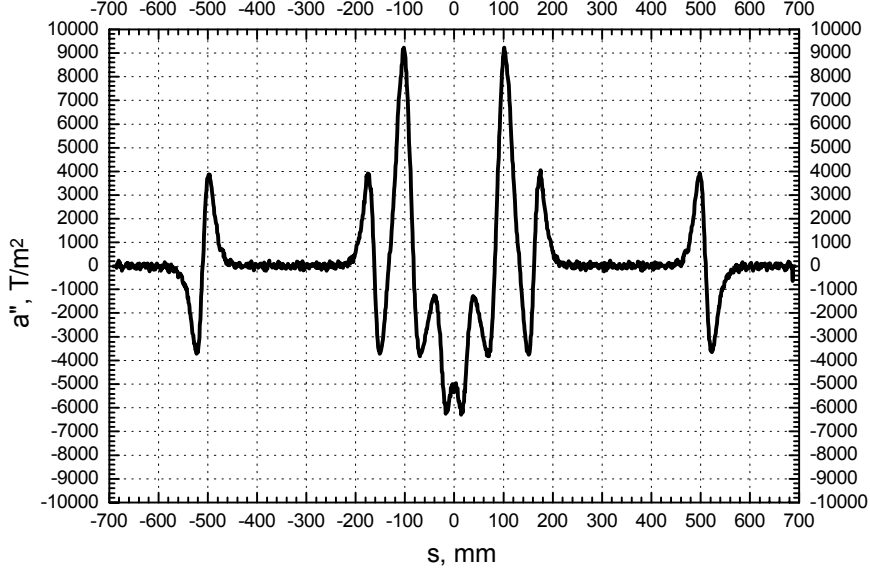

Fig.7. Measured edge field pseudo-sextupole profile

We have to take into account the focusing coming from the sextupolar field of the wiggler (Fig.8). This focusing is caused by the orbit deviation and the presence of the nonlinear fields in the wiggler. The focusing strength is $\int_{-L_w/2}^{L_w/2} k_{zs} ds = -\frac{1}{B\rho} \int_{-L_w/2}^{L_w/2} G_s ds = 0.706$ m$^{-1}$. In this case the integrated vertical and horizontal coefficients of the central pole are $I_{zw} = \int_{-L_w/2}^{L_w/2} k_y ds = 0.776 + 0.706 = 1.482$ m$^{-1}$, $I_{xw} = \int_{-L_w/2}^{L_w/2} \left( \frac{1}{\rho^2(s)} - k_z(s) \right) ds = 0.776 - 1.482 = -0.704$ m$^{-1}$. For BM2 (Fig.6) we have to recalculate the vertical and horizontal focusing strengths: $k_z = \frac{I_{zw}}{2L_2} = 18.52$ m$^{-2}$, $k_x = \frac{I_{xw}}{2L_2} = -8.804$ m$^{-2}$.

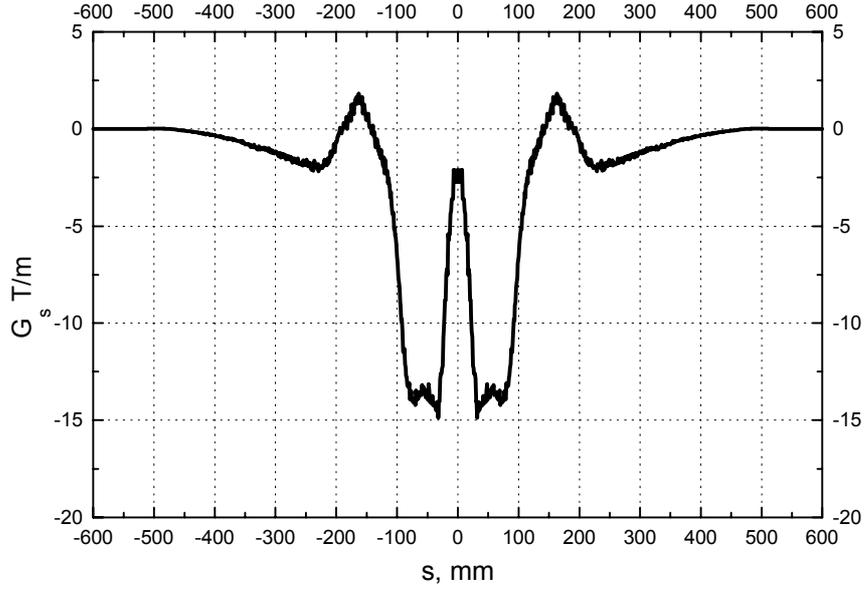

Fig.8. Focusing coming from the sextupolar field of the wiggler

Taking into account all above considerations, we propose the following model of the magnetic field of the wiggler for the computer simulations (Table 1).

Table 1. 10 T wiggler model for computer simulations

| Item | | Length, m | Bending radius, m | Focusing strength |
|---|---|---|---|---|
| Side pole | | 0.35 | 2.118 | $K_1$=-0.22 m$^{-2}$ |
| Drift | | 0.049 | | |
| Sextupole | | 0 | | $K_2$=-12.95 m$^{-3}$ |
| Central pole | Side magnet BM2 | 0.0406 | 0.45 | $K_1$=-9.55/-18.52 m$^{-2}$ * |
| | Central magnet BM1 | 0.06 | 0.4 | $K_1$=0 |
| | Side magnet BM2 | 0.0406 | 0.45 | $K_1$=-9.55/-18.52 m$^{-2}$ * |
| Sextupole | | 0 | | $K_2$=-12.95 m$^{-3}$ |
| Drift | | 0.049 | | |
| Side pole | | 0.35 | 2.118 | $K_1$=-0.22 m$^{-2}$ |

\* focusing strength without/with focusing coming from the sextupolar field of the wiggler.

## 3. The influence of the insertion devices on the DELSY beam dynamics

*3.1. DELSY lattice and basic parameters*

A layout with four straight sections was chosen for the DELSY storage ring. The periodicity of the ring is 2. Every quadrant consists of the MBA structure: two halves of straight sections and two periodic cells. The periodic cell consists of two dipoles and three quadrupoles. The matching cell contains two dipoles and provides zero dispersion in the straight section. A doublet adjusts the particular values of the beta functions in the straight sections. The basic machine parameters are given in Table 2.

Table 2. Basic parameters of the DELSY ring

| Circumference, m | 136.04 |
|---|---|
| Bending radius, m | 3.3 |
| Energy, GeV | 1.2 |
| Injection energy, GeV | 0.8 |
| Momentum compaction factor | $5.03 \cdot 10^{-3}$ |
| Chromaticity (hor./vert.) | -22.2/-12.6 |
| Stored current, mA | 300 |
| Horizontal emittance, nm·rad | 11.4 |
| Betatron tunes (hor./vert.) | 9.44/3.42 |
| RF frequency, MHz | 476 |
| Harmonic number | 216 |
| Bunch length, mm | 8.67 |
| Energy loss per turn, keV | 55.7 |
| Natural energy spread | $5.56 \cdot 10^{-4}$ |

The beta functions in a very strong wiggler must be small enough to avoid an increase in the emittance and to minimise the optics distortions with the wiggler on. In our case $\beta_x=1.05$ m and $\beta_y=2.80$ m. The vertical beta function in the centre of the undulator must be small to provide the tolerable lifetime limited by the residual gas scattering. It was taken to be $\beta_x=14.55$ m and $\beta_y=0.98$ m (Fig.9). In another "undulator" quadrant the injection septum is placed. Two injection kickers separated in $9\pi$ betatron phase advance are located in the "wiggler" quadrants.

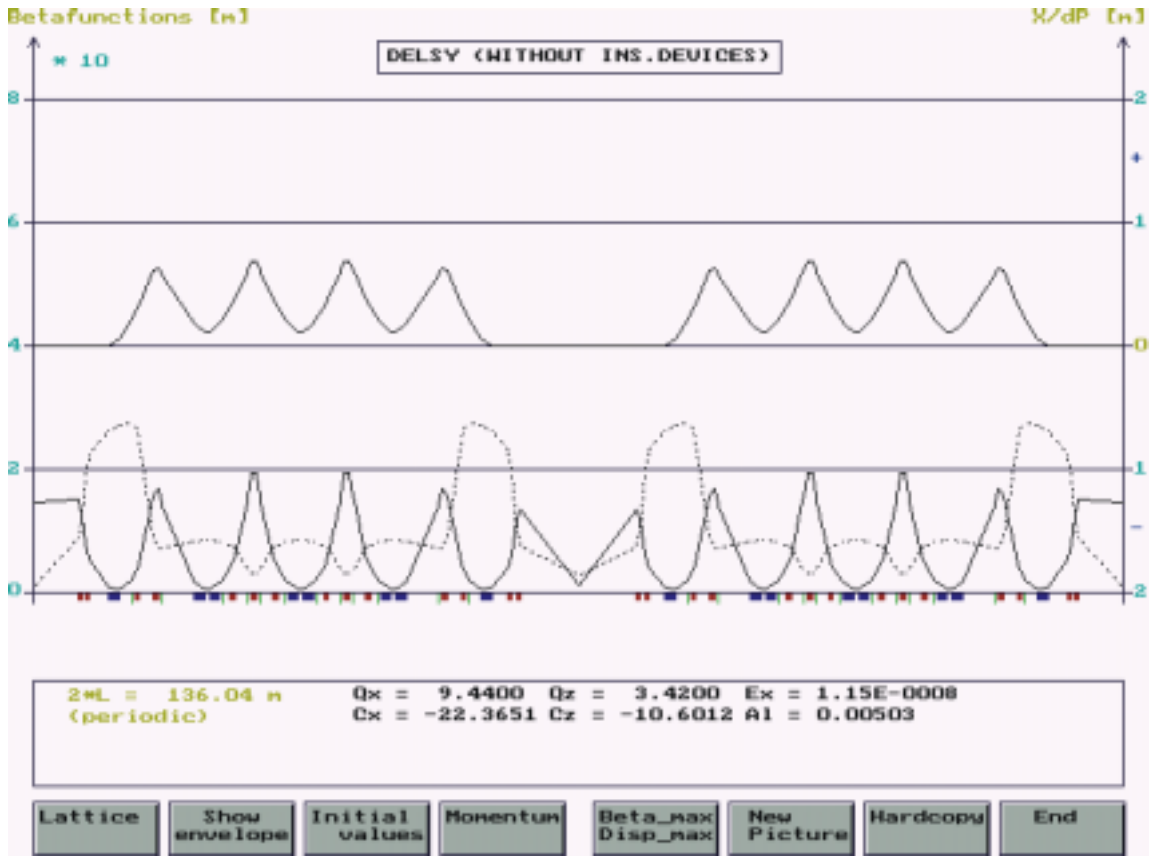

Fig.9. Lattice functions for the half of the ring

The injection energy for DELSY is 0.8 GeV, while operation is at 1.2 GeV. This imposes strong constraints on the dynamic aperture of DELSY. The solution with two sextupole families that fits requirements was found. First, phase advances of the periodic cell were chosen to provide a small emittance with tolerable natural chromaticities. Then the positions of the sextupoles in the matching cells were adjusted. Since the wiggler insertion breaks the ring symmetry from 2 to 1 and the number of resonances in the vicinity of the working point increases drastically, non-linear beam dynamics and dynamic aperture shrinking may arise to become a serious problem. Hence, usual fast determination of the dynamic aperture border now seems not enough. To save computing time, most codes define the dynamic aperture "from outside", when a particle starts from the surely unstable region and then step-by-step counts down the initial motion amplitude until the stable region is found. An evident drawback of this method is that if the particle runs across the stable island of a high-order resonance, the wrong dynamic aperture will be found. To avoid this mistake we additionally apply a different technique of a surviving plot "from inside". The particle scans the amplitude space starting from the coordinate origin and the number of survived revolutions is plotted as a function of the initial amplitudes $(A_x, A_z)$. In our case the maximum number of turns is 1000. This algorithm is rather time-consuming but yields more reliable and detailed information.

The on-energy dynamic aperture of the machine caused by the chromatic sextupoles (the wiggler is switched off) is shown in Fig.10. The dynamic aperture is plotted for the initial lattice azimuth with the natural chromaticity corrected to 0 in both planes. Calculations were made with the OPA computer code [6].

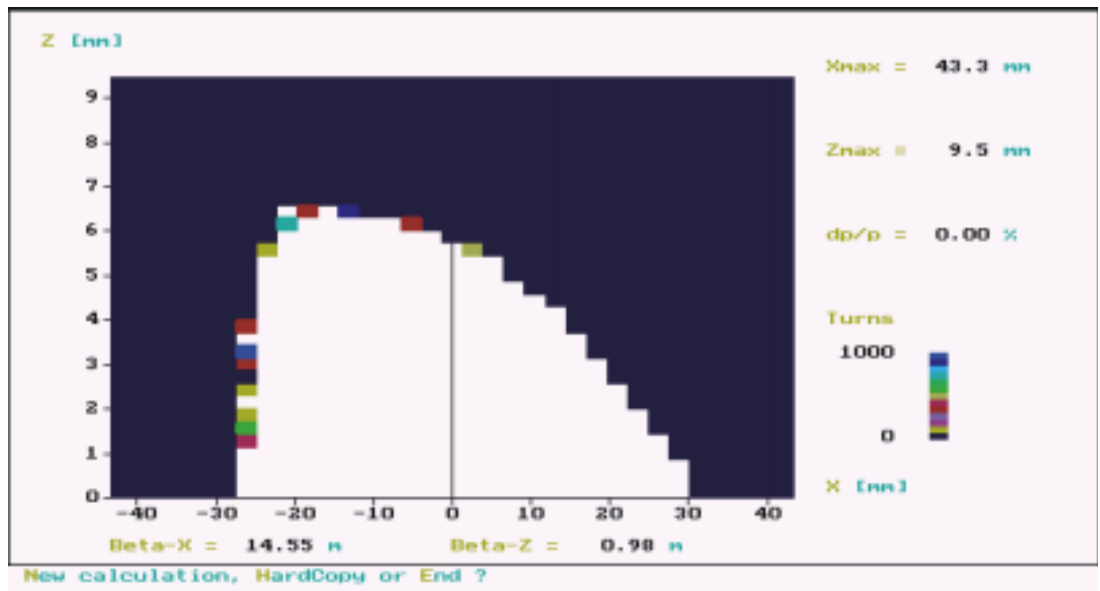

Fig.10. Survival plot of the bare DELSY ring
(white square is the stable particle, the other is unstable)

*3.2. The influence of the 10 T wiggler on the linear optics and dynamic aperture*

Inserting the 10 T wiggler in the way described in the previous section produces an unstable lattice, so the ring cannot operate without compensation for the wiggler influence. The following procedure was used to minimise great distortion of the linear optics. First, the strengths of two quadrupoles in the doublet matching the wiggler section were modified to maintain constraints $(\alpha_x=\alpha_y=0)$ with the wiggler on and off. This prevents the beating of the beta functions everywhere outside the wiggler section. After this the machine tunes are changed significantly. To bring them back and to maintain the required beta functions in the straight sections, a global matching procedure involving all quadrupole families were used. This procedure was made for two models of the wiggler's magnetic field: for the model without focusing coming from the sextupolar field of the wiggler in the side magnet of the central pole and for the model with them (see Table 1), as we have to define the requirements to the wiggler.

For the first model the deviation of the beta functions is less than 7% (Fig.11), the emittance increased by a factor of 1.9 and the natural energy spread increased by a factor of 1.72 (Table 3).

Table 3. Basic parameters of the ring without/with the wiggler for the first model

|  | Wiggler is off | Wiggler is on |
|---|---|---|
| Momentum compaction factor | $5.03 \cdot 10^{-3}$ | $4.98 \cdot 10^{-3}$ |
| Chromaticity (hor./vert.) | -22.2/-12.6 | -22.0/-11.2 |
| Horizontal emittance, nm·rad | 11.4 | 21.3 |
| Bunch length, mm | 8.67 | 14.95 |
| Energy loss per turn, keV | 55.7 | 82.9 |
| Natural energy spread | $5.56 \cdot 10^{-4}$ | $9.55 \cdot 10^{-4}$ |
| Maximum gradient in the quadrupoles, T/m | 19.1 | 19.58 |
| Maximum gradient in the sextupoles, $T/m^2$ | 88.6 | 93.0 |

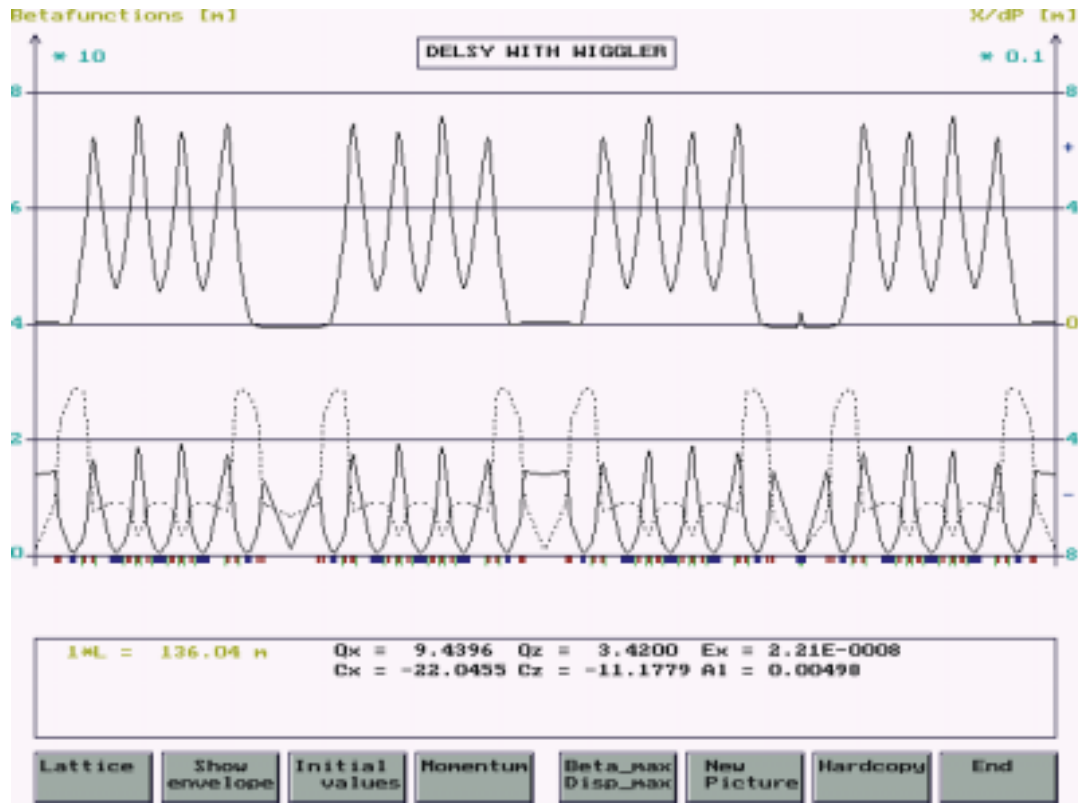

Fig.11. Lattice functions for the DELSY ring (wiggler is on), first model

In spite of the increased number of high-order resonances near the stable motion boundary, the dynamic aperture still seems sufficient for the reliable machine operation (Fig.12).

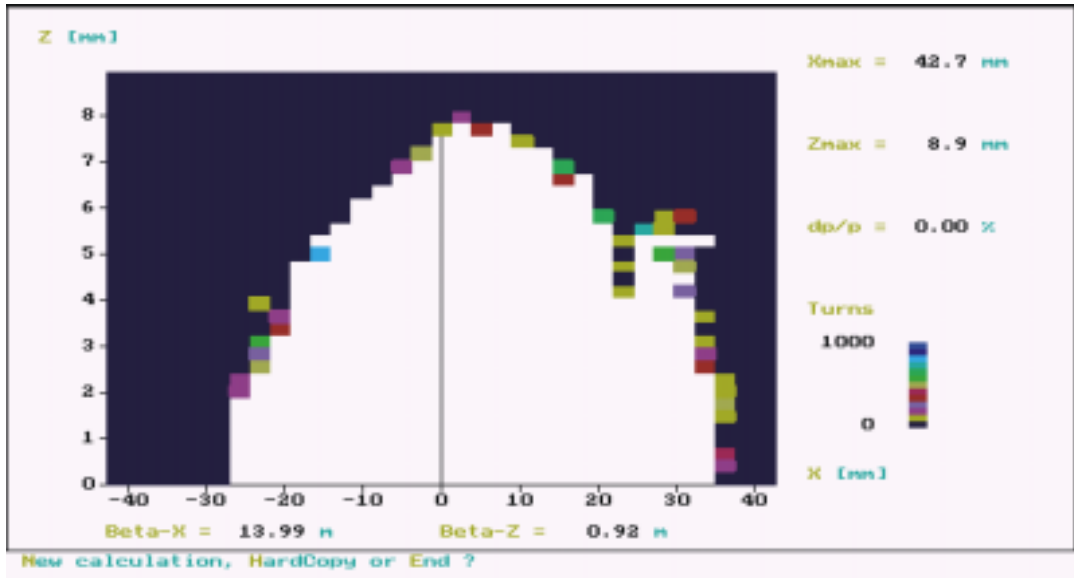

Fig.12 Dynamic aperture after compensation of the 10 T wiggler influence (first model)

For the second model the correction of the wiggler influence on the linear optics is more difficult: the gradients of the quadrupoles are bigger that in the previous model, additional power supply is needed for the quadrupoles placed in the matching cell near the wiggler. The maximum of the horizontal beta function increases to 78 m (Fig.13), the emittance increases to 39 nm, chromaticity is bigger than in the previous model too and equals -28.22/-12.93.

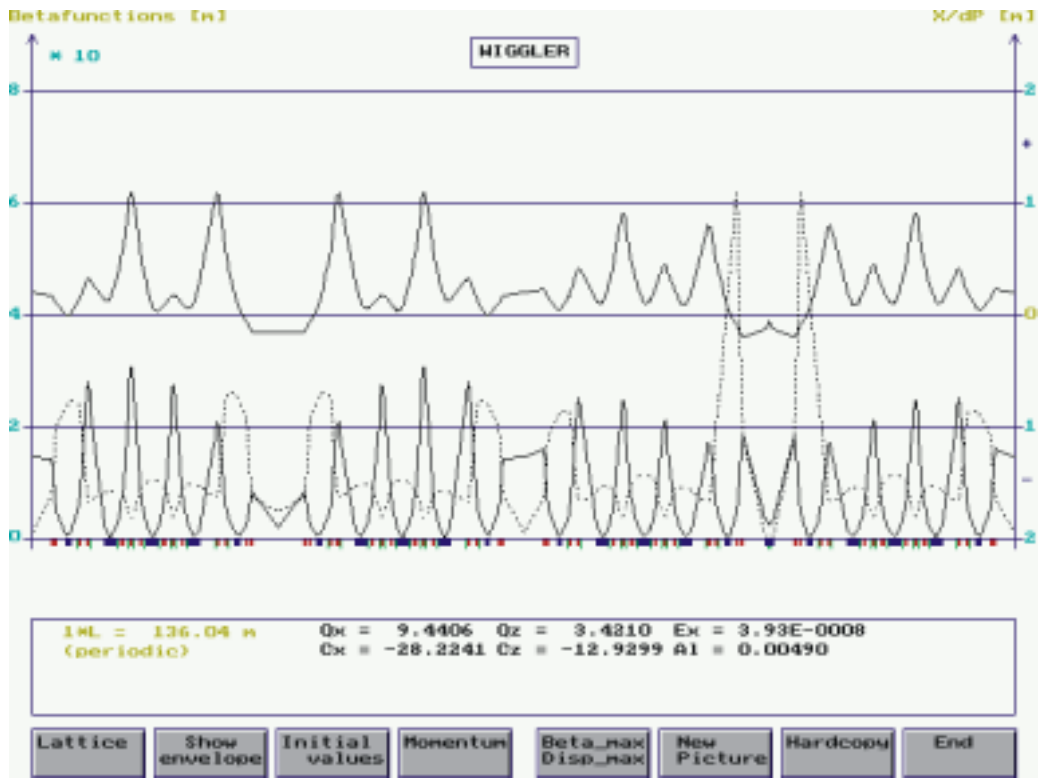

Fig.13. Lattice functions for the DELSY ring (wiggler is on), second model

For the second model of the wiggler's magnetic field the shrinking of the dynamic aperture for the regime with the wiggler on is more essential (Fig.14).

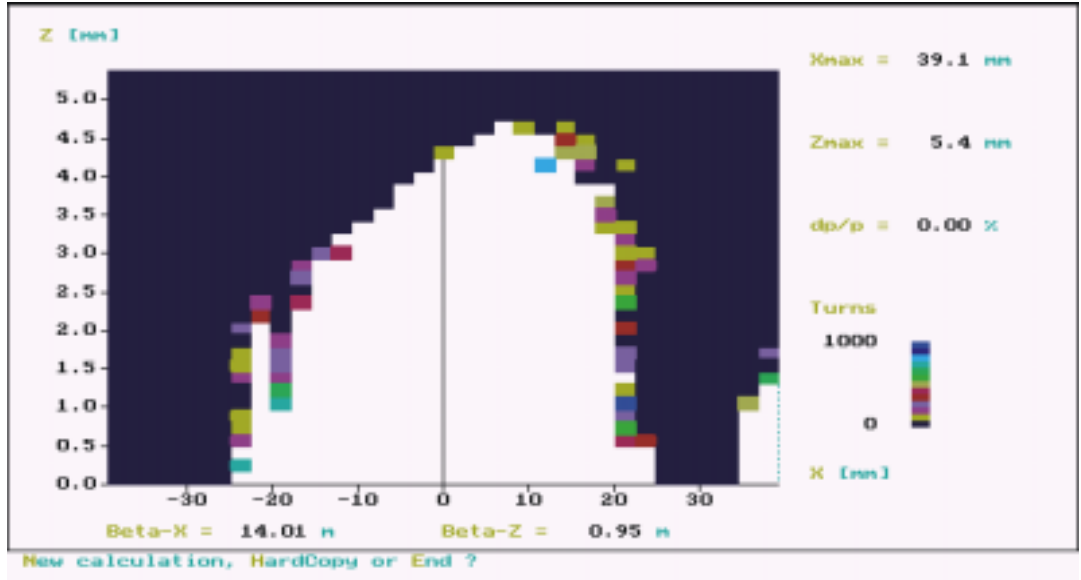

Fig.14. Dynamic aperture after compensation of the 10 T wiggler influence for the second model of the wiggler's magnetic field

The main conclusion is that the wiggler has to be manufactured in the way to reduce the focusing coming from the sextupolar field. One of the ways to do this is as follows: the beam is declined by a pair of dipole correctors (upstream and downstream the wiggler) so as to keep always the radiation source point in the middle of the central pole with zero displacement. In this case particles travel in the high-field region near the wiggler axis and we can expect that the influence of transverse nonlinearity will be significantly reduced. The other advantage is the fixed geometry of the synchrotron radiation light for different field levels.

*3.3. The influence of the undulator on the linear optics and dynamic aperture*
Within the computer code OPA we can describe the undulator as an individual element. The effect of the undulator (0.75 T, 150 periods of 1.5 cm) on the machine optics is much smaller than the effect of the wiggler (Table 4). The deviation of the beta functions for the machine with the undulator on is less than 1 % (Fig.15), the emittance increases to 11.8 nm.

Table 4. Basic parameters of the ring without/with undulator

|  | Undulator is off | Undulator is on |
|---|---|---|
| Momentum compaction factor | $5.03 \cdot 10^{-3}$ | $5.02 \cdot 10^{-3}$ |
| Chromaticity (hor./vert.) | -22.2/-12.6 | -22.9/-12.2 |
| Horizontal emittance, nm·rad | 11.4 | 11.8 |
| Bunch length, mm | 8.67 | 8.7 |
| Energy loss per turn, keV | 55.7 | 570.5 |
| Natural energy spread | $5.56 \cdot 10^{-4}$ | $5.54 \cdot 10^{-4}$ |
| Maximum gradient in the quadrupoles, T/m | 19.1 | 19.32 |
| Maximum gradient in the sextupoles, T/m$^2$ | 88.6 | 89.1 |

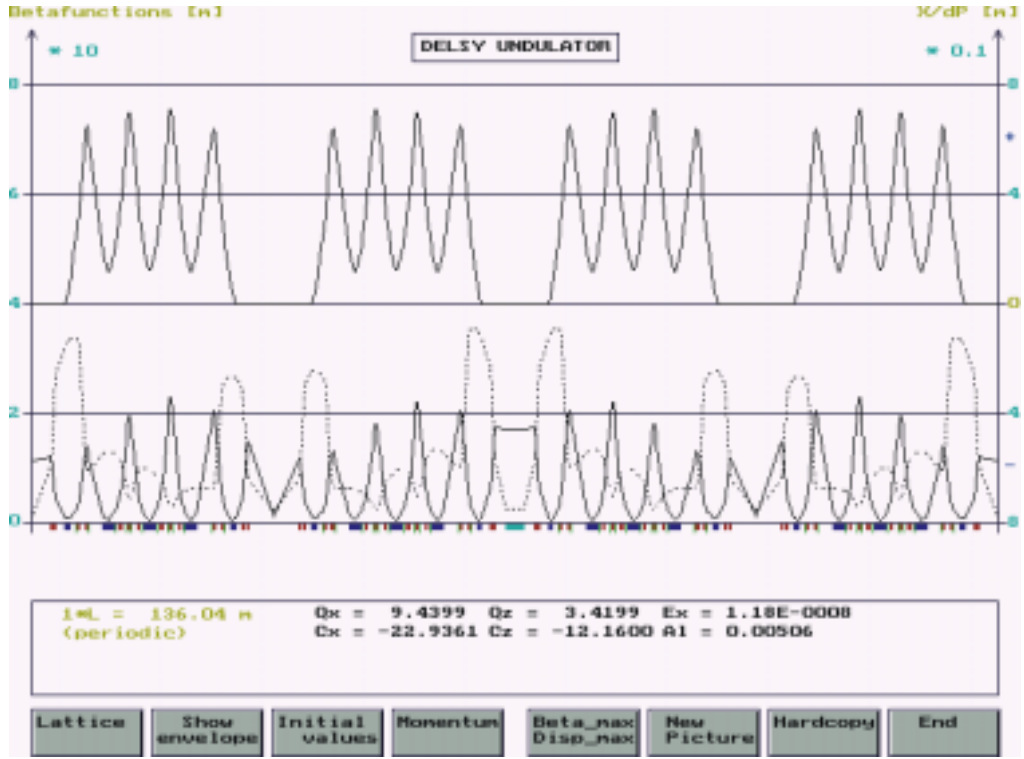

Fig.15. Lattice functions for the regime with the undulator on

*3.4. The influence of the both insertion devices on the DELSY beam dynamics*

We calculated the regime with both insertion devices on only for the first model of the wiggler's magnetic field. The basic parameters of this regime are presented in Table 5. Due to the additional quantum excitation the horizontal emittance has grown up by a factor of 1.78 in comparison with the initial regime without the insertion devices. This reduces the light source brightness and increased vertical betatron function (Fig.16).

Table 5. Basic parameters of the ring for the regime with the wiggler and the undulator on

| Momentum compaction factor | $5.03 \cdot 10^{-3}$ |
|---|---|
| Chromaticity (hor./vert.) | -22.3 /-11.4 |
| Horizontal emittance, nm·rad | 20.3 |
| Bunch length, mm | 14.91 |
| Energy loss per turn, keV | 824 |
| Natural energy spread | $9.48 \cdot 10^{-4}$ |
| Maximum gradient in the quadrupoles, T/m | 19.62 |
| Maximum gradient in the sextupoles, T/m$^2$ | 91.96 |

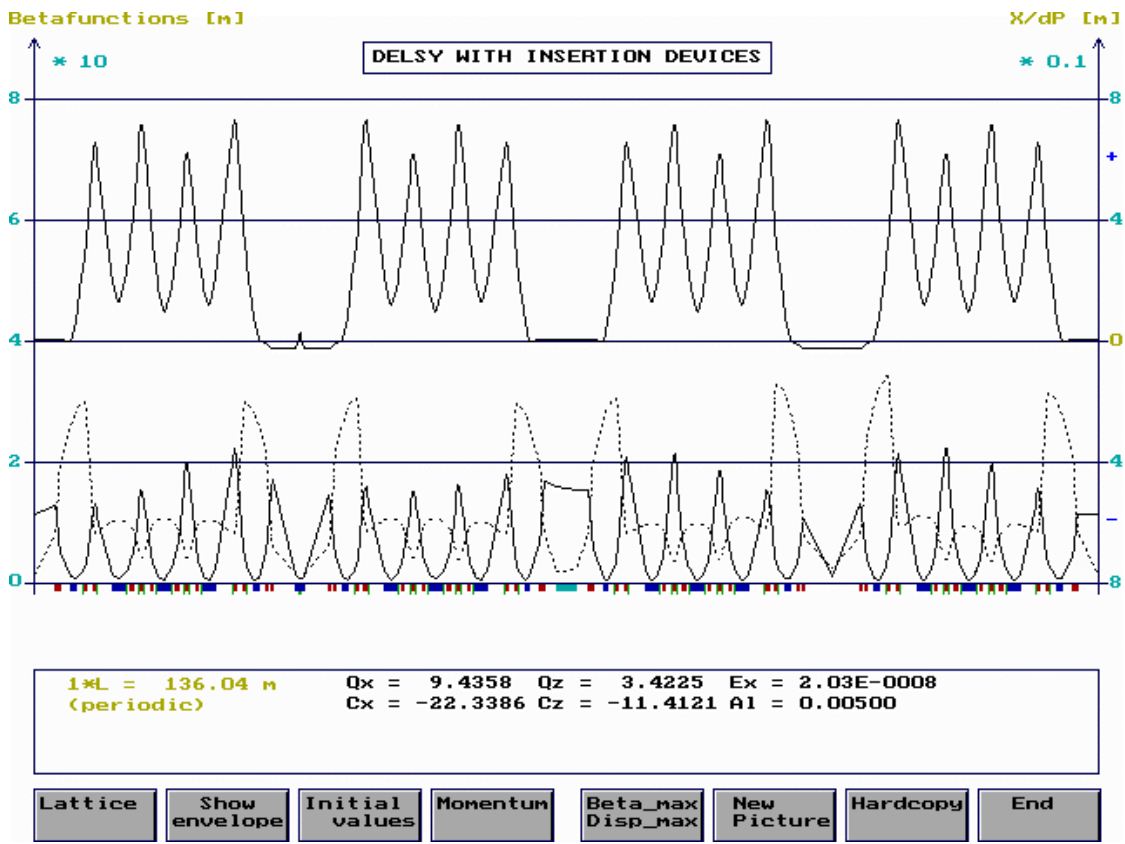

Fig.16. Lattice functions for the regime with both insertion devices

The dynamic aperture is plotted in Fig.17. The analysis shows that the dynamic aperture is reduced mainly due to the breaking of the ring symmetry.

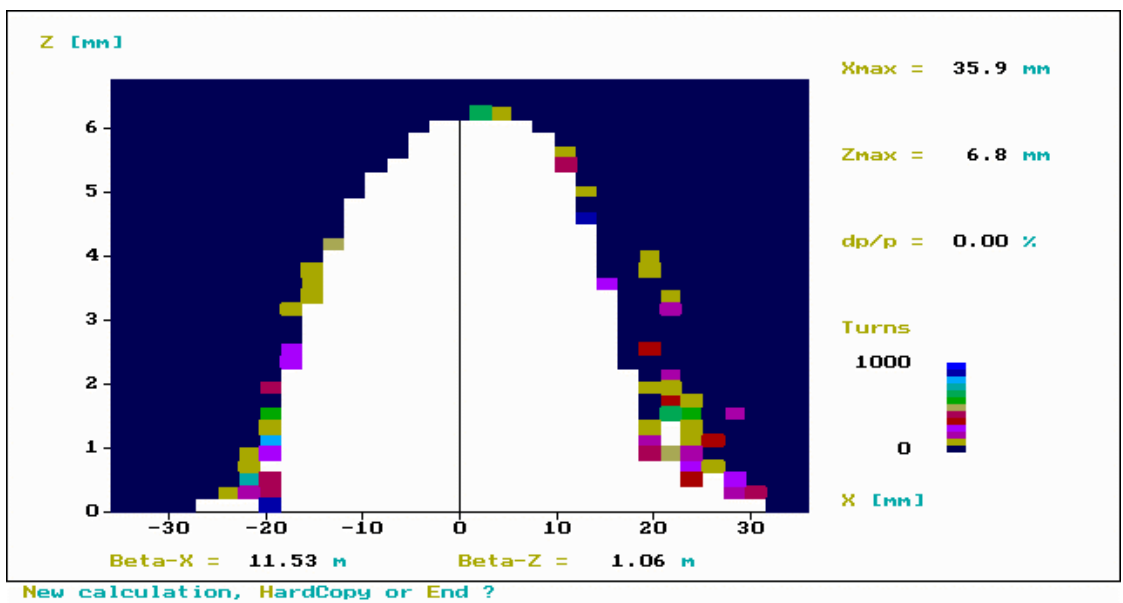

Fig.17. Dynamic aperture after compensation of the wiggler and undulator influences

## 4. Conclusions

The magnetic field model is elaborated for the 10 T wiggler. The result of the wiggler influence on the DELSY beam dynamics is valid only for the wiggler model described above. This model, based on magnetic field expansion around central orbit in the wiggler, represent its main properties including focusing. The following conclusions can be drawn from the present study:

1) the 10 T wiggler significantly influences the storage ring dynamics;
2) this influence can be recovered by applying local and global linear optics correction;
3) when the linear lattice is cured well, the reduction of dynamic aperture with the wiggler and undulator on is not big;
4) the emittance increases by a factor of 1.78 for the regime when both insertion devices are switched on;
5) the main requirement for the wiggler design is to decrease focusing coming from the sextupolar field of the wiggler.

**Acknowledgements**

We would like to thank Dr.N.Mezentsev for the help and for the placing at our disposal the magnetic measurement data of the 10 T wiggler.